\documentclass[12pt,fleqn]{article} 
\usepackage{graphics} 
\usepackage{cite}
\usepackage{amsfonts} 
\newcommand{\gsim}{\stackrel{\scriptstyle >}{{ }_{\sim}}}

\newcommand{\mb}{\ensuremath{m_b}}
\newcommand{\mt}{\ensuremath{m_t}}
\newcommand{\Dmb}[1][]{\ensuremath{\Delta\mb^{#1}}}

\newcommand{\Dmf}[1][]{\ensuremath{\Delta m_f^{#1}}}
\newcommand{\tb}[1][]{\ensuremath{\tan^{#1}\!\beta}}

\newcommand{\mg}{\ensuremath{M_{\tilde{g}}}}
\newcommand{\msb}[1]{\ensuremath{M_{\tilde{b}_{#1}}}}
\newcommand{\mst}[1]{\ensuremath{M_{\tilde{t}_{#1}}}}
\newcommand{\mstau}[1]{\ensuremath{M_{\tilde{\tau}_{#1}}}}

\newcommand{\TeV}{\ensuremath{\,{\rm TeV}}}
\newcommand{\GeV}{\ensuremath{\,{\rm GeV}}}
\newcommand{\rsm}{\ensuremath{R^{\rm SM}}}
\newcommand{\rmssm}{\ensuremath{R^{\rm MSSM}}}
\newcommand{\brHbb}{\ensuremath{BR(H\to b\bar{b})}}
\newcommand{\brHtt}{\ensuremath{BR(H\to \tau^+\tau^-)}}
\newcommand{\mtau}{\ensuremath{m_\tau}}

\newcommand{\ta}{\ensuremath{\tan\alpha}}

\newcommand{\ma}{\ensuremath{M_{A^0}}}

\newcommand{\Dmtau}{\ensuremath{\Delta m_\tau}}
\begin{document} 
\hfill{} 
\begin{tabular}{l} 
MPI-PhT/2003-28\\ 
PSI-PR-03-08\\
LC-TH-2003-043\\
hep-ph/0307012\\
\vspace*{0.4cm}
\end{tabular} 
\renewcommand{\thefootnote}{\fnsymbol{footnote}}
\begin{center}
\textbf{\large Some results on the distinction of Higgs boson models}{\footnote{Based on talk given at Second Workshop of the Extended 
ECFA/DESY Study, Saint Malo, April 12-15, 2002}}
\vspace*{0.7cm}\\
{\par\centering 
Jaume Guasch$^{\,\small{a}}$, Wolfgang Hollik$^{\,\small{b}}$, 
{\underline{Siannah Pe{\~n}aranda}}$^{\,\small{b}}$~{\footnote{
electronic address: jaume.guasch@psi.ch, 
hollik@mppmu.mpg.de,
siannah@mppmu.mpg.de}}
\par} 
\vspace*{0.3cm}
{\par\centering 
\textit{$^{\,\small{a}}$ Paul Scherrer Institut, CH-5232 Villigen PSI, Switzerland}\\
\textit{$^{\,\small{b}}$ Max-Planck-Institut f\"{u}r Physik,  F\"{o}hringer Ring 6,
    D-80805 M\"{u}nchen, Germany }}
\end{center}
\vspace*{0.7cm}
{\par\centering\textbf{\large Abstract}\vspace*{0.5cm}\\ 
\par} 
\noindent 
{\small{We present results on the analysis of the ratio of branching
    ratios 
$R=\brHbb/\brHtt$ of Higgs boson decays as
a discriminant quantity between supersymmetric  and
non-super\-symmetric models.
A detailed analysis in the
effective Lagrangian approach shows how one could discriminate between
models at the Large Hadron Collider and the $e^+e^-$
Linear Collider at $500\GeV$ center of mass energy.}}
\renewcommand{\thefootnote}{\arabic{footnote}}
\setcounter{footnote}{0}

\vspace*{0.6cm}
The search for a Higgs boson is nowadays the main objective of High
Energy Physics experiments. Even if a neutral scalar boson is
discovered in present or future colliders, the question 
will still be open: whether it is the Higgs particle of the
minimal Standard Model (SM) 
or whether there is an extended Higgs structure beyond the SM.
We approach this question by investigating the neutral Higgs sector
of various types of models. In particular, we consider 
the ratio of
branching ratios of a neutral Higgs boson $H$,
\begin{equation}
  \label{eq:Rdef}
  R=\frac{\brHbb}{\brHtt}\,\,,
\end{equation}
and we  analyze in detail the Yukawa-coupling effects
and their phenomenological consequences~\cite{ours}. 
This ratio receives large renormalization-scheme independent radiative
corrections in supersymmetric (SUSY) models at large $\tan \beta$,  
which are absent in the SM or Two-Higgs-doublet models (THDM). These corrections are
insensitive to the supersymmetric mass scale. We consider in our
analysis the effective Lagrangian approach by 
 relating the quark mass to the Yukawa coupling via $\Dmf$,
the non-decoupling quantity that encodes the leading radiative
corrections,
\begin{equation}
  \label{eq:deffhb}
  h_f=\frac{m_f(Q)}{v_1} \frac{1}{1+\Dmf}=
      \frac{m_f(Q)}{v \cos\beta}\frac{1}{1+\Dmf}, \quad \quad
    v= (v_1^2+v_2^2)^{1/2} \, .
\end{equation}
Here the resummation of all possible $\tb$
enhanced corrections of the type $(\alpha_{(s)} \tb)^n$~\cite{eff} are
included. The leading part of the (potentially) non-decoupling
contributions proportional to $A_b$ can be absorbed in the definition of
the effective Yukawa coupling at low energies and only subleading
effects survive~\cite{Guasch-Spira}. Therefore,
expression~(\ref{eq:deffhb}) contains all leading potentially large
radiative effects.

Notice that the ratio~(\ref{eq:Rdef}) is very interesting from both 
the experimental and the theoretical side.
It is a clean observable, measurable in a
counting experiment, with only small 
systematic errors since most of them cancel in the ratio. The only
surviving systematic effect results from the efficiency of
$\tau$- and $b$-tagging. 
From the theoretical side,
it is  independent of the production
mechanism of the decaying neutral Higgs boson and of the total 
width; hence,  new-physics effects affecting the production cross-section do
not appear in the ratio~(\ref{eq:Rdef}). For the same reason, 
this observable is insensitive to unknown
high order QCD corrections to Higgs boson production. 
Besides, since this ratio only depends on the ratio of 
the masses, there is no other parameter (e.g.\  \tb) that could absorb
the large quantum corrections. 

The normalized value of the ratio of the Higgs boson decay rates into
$b-$quarks and $\tau-$leptons~(\ref{eq:Rdef}) 
to the $\rsm$ expectation is used to
distinguish a general THDM from the Minimal Supersymmetric Standard
Model (MSSM). 
We analyze the deviation of this ratio from the SM value, 
caused by the SUSY radiative corrections, for each  
of the MSSM neutral Higgs bosons $\phi=h,H,A$. This normalized value 
is a function depending only on $\tb$, $\ta$, $\Dmb$ and
$\Dmtau$,  and encoding all the genuine SUSY corrections. 
The explicit form of $\Dmb$ and $\Dmtau$ at the one-loop level can be 
obtained approximately 
by computing the supersymmetric loop diagrams at zero external momentum
($M_{SUSY} \gg m_b\,,m_\tau$)~\cite{ours}. 
These two quantities are independent of the SUSY
mass scale $M_{SUSY}$ since they only depend on $\tb$ and 
the ratio $A_t/M_{SUSY}$~\cite{ours}~{\footnote{See also ~\cite{eff,CMW}.}}.

The genuine SUSY corrections present sizeable differences 
between the $\phi bb$ and $\phi\tau\tau$
couplings, even in the case of similar squark and slepton spectra:
the SUSY-QCD corrections mediated by gluinos 
are only present in $\Dmb$, 
yielding the by
far dominant contribution to the normalized value of $R$;
there exists a contribution from the
chargino sector to $\Dmb$ resulting from mixing in the stop sector, 
whereas a corresponding term is not present 
in $\Dmtau$  due to the absence sneutrino mixing;
the contribution from the $\tilde{B}$ loops is
different in both cases because of the different hypercharges.

As for the case of the lightest MSSM CP-even Higgs boson, $h^{0}$, 
the ratio $R$ defined in~(\ref{eq:Rdef}), written in terms of
the non-decoupling quantities $\Dmb$ and $\Dmtau$ and normalized to the
SM value, reads
\begin{equation}
\label{eq:Rh0}
\frac{\rmssm(h)}{\rsm}=\frac{(1 + \Dmtau)^2\,(-\cot\alpha \Dmb + \tan \beta)^2}
{(1 + \Dmb)^2\,(-\cot\alpha \Dmtau + \tan \beta)^2}\,.
\end{equation}
\begin{figure}[t]
\begin{center}
\begin{tabular}{cc}
\resizebox{6.1cm}{!}{\includegraphics{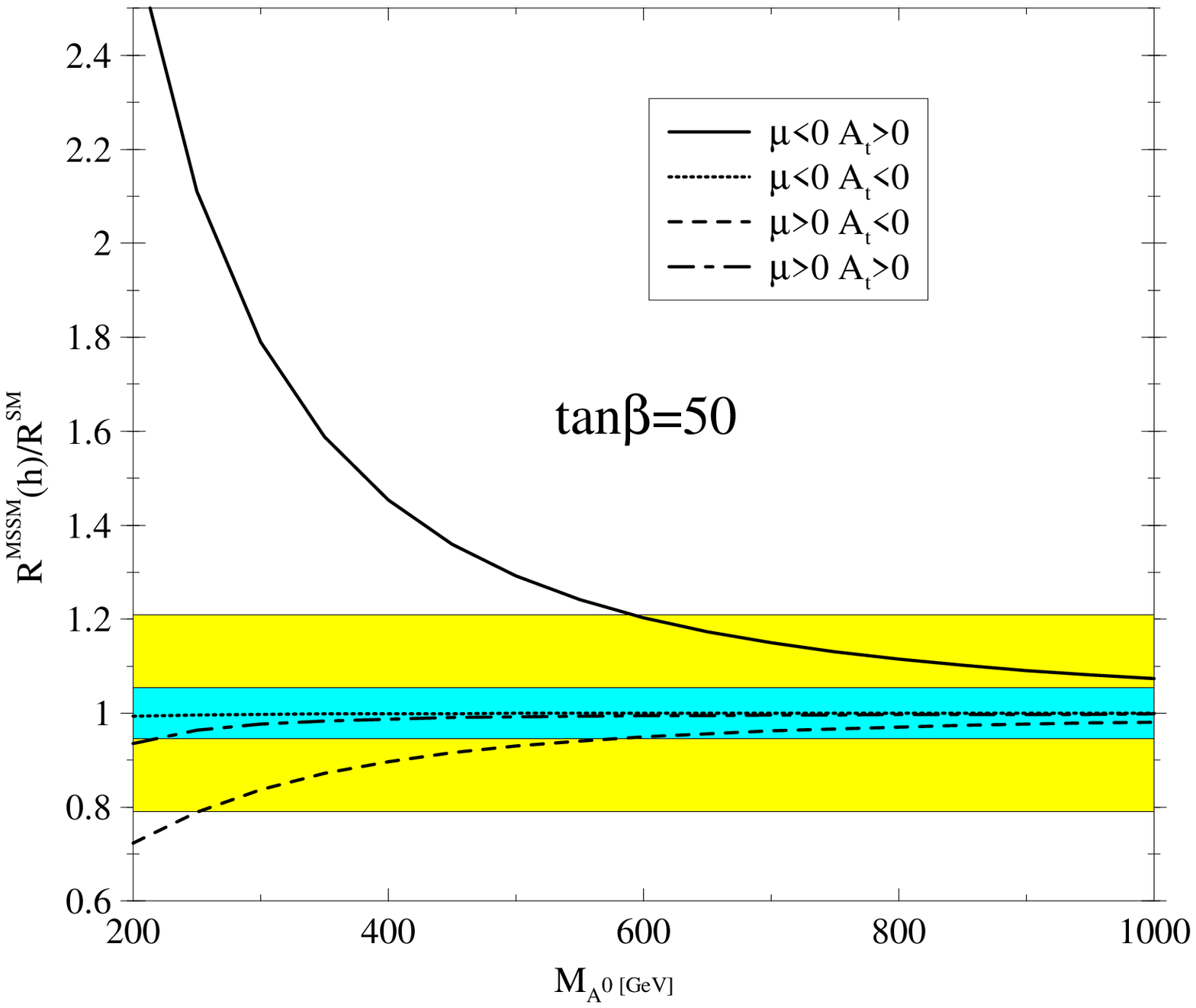}}&
\resizebox{6cm}{!}{\includegraphics{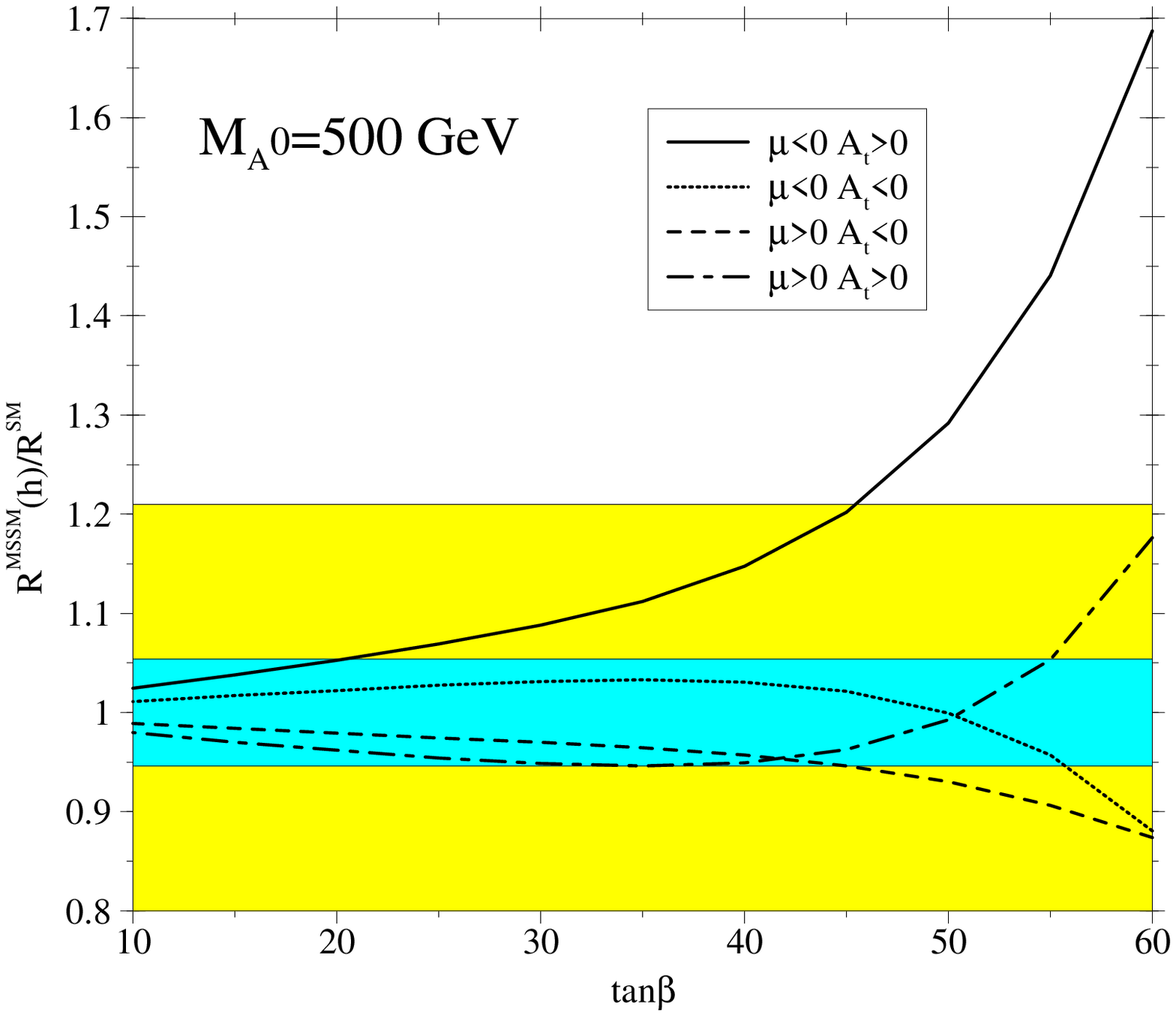}}\\
(a)&(b)
\end{tabular}
\end{center}\vspace*{-0.4cm}
\caption{Deviation of $\rmssm(h)$ with respect to the SM value, as a
  function of \textbf{a)} \ma, and \textbf{b)} \tb, for various choices
  of the SUSY parameters. The light-shaded region shows the $\pm21\%$
  deviation with respect to the SM, and the dark-shaded one the
  $\pm5.4\%$.}
\label{fig:rmssmh}
\end{figure}

Large deviations from $R=1$ are expected in the $\rmssm$ for several
MSSM parameter combinations as shown in Fig.~\ref{fig:rmssmh}. 
The SUSY spectrum has been taken to be around $1.5\TeV$, namely, 
$\mg=\msb1=\mst1=\mstau1=M_2=|\mu|=A_b=A_\tau=|A_t|=1.5\TeV\,\,,$
and we assume the usual GUT relation $M_1=5/3 M_2 s^2_W/c^2_W$ and maximal
mixing in the $\tilde{b}$  and $\tilde{\tau}$ sector, $\theta=\pm\pi/4$. 
The rest of
the parameters are fixed by the $SU(2)_L$ symmetry. As for the SM
parameters we have chosen 
$\mt=175\GeV$, $\mb=4.62\GeV$, $\mtau=1.777\GeV$~\cite{PDG}.  
The CP-even
mixing angle is computed including the leading 
corrections up to two-loop order by means of the  
program \textit{FeynHiggsFast}~\cite{FeynHiggsFast}.

\begin{figure}[t]
\begin{center}
\begin{tabular}{cc}
&\\
\resizebox{6.1cm}{!}{\includegraphics{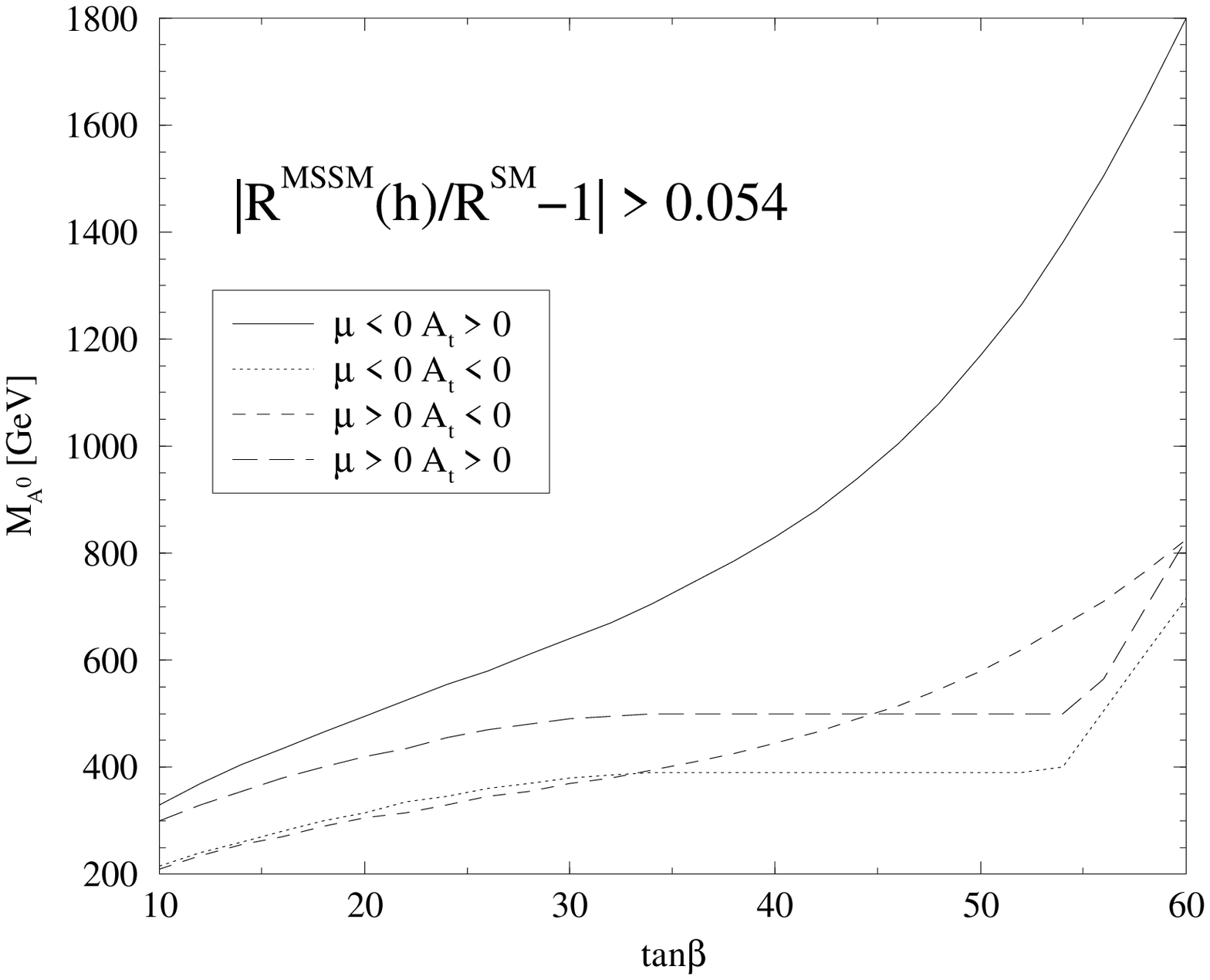}}&
\resizebox{6cm}{!}{\includegraphics{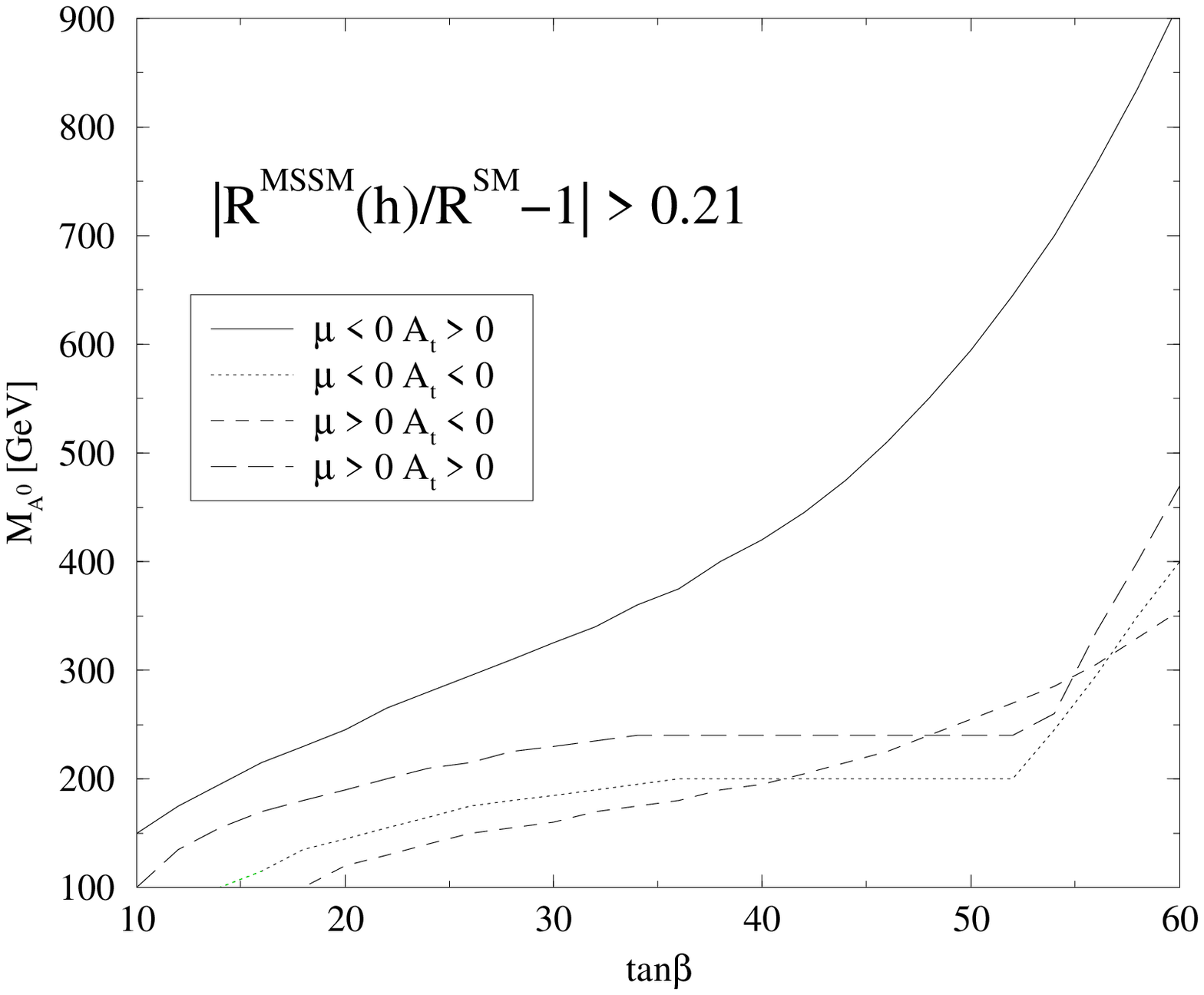}}\\
(a)&(b)
\end{tabular}
\end{center}\vspace*{-0.5cm}
\caption{Sensitivity regions on $\rmssm/\rsm$ with \textbf{a)} 5.4\%
  uncertainty in the measurement; \textbf{b)} 21\% uncertainty.}
\label{fig:exclregion}
\end{figure}
The decoupling behaviour with \ma\  becomes apparent in
Fig.~\ref{fig:rmssmh}a.
In some favorable cases, i.e. small \ma, large \tb, $\mu<0$
and $A_t>0$,  the 
ratio (\ref{eq:Rh0}) can be as large as two. 
Clearly, a moderate-precision
measurement of this quantity would give clear signs of a Higgs boson
belonging to a SUSY model. For the LHC we estimate that this quantity
can be measured to a $21\%$ accuracy. 
By looking at the associate $WW$-fusion Higgs boson production
$qq\rightarrow W^*W^*\to H$, the 
$BR(H\to \tau^+ \tau^-)/BR(H\to \gamma \gamma)$ is measurable with an accuracy
of order $15\%$~\cite{Zeppenfeld}. On the other hand, for the associated
Higgs-boson production with a top quark
$(pp\to t\bar t H)$ the ratio $BR(H\to b\bar{b})/BR(H\to\gamma\gamma)$
can be performed with a similar precision~\cite{Gianotti}. From these
two independent 
measurements one determines $R$ with the error quoted above. If one were
able to make both measurements using the same Higgs-boson production
process, the error might be decreased.
The $\pm21\%$ deviation region is marked as a light-shaded region in the
figures. As for a future $e^+e^-$ Linear Collider (LC) running at $500\GeV$
center-of-mass energy, the simulation shows that the ratio of the 
effective Yukawa  couplings, $h_b/h_\tau(\equiv\sqrt{R})$, 
can be measured with an
accuracy of $2.7\%$~\cite{TESLATDR}. The corresponding band of $\pm
5.4\%$ accuracy in~(\ref{eq:Rh0}) is shown as a dark-shaded region. 

\begin{figure}[t]
\centerline{\resizebox{6cm}{!}{\includegraphics{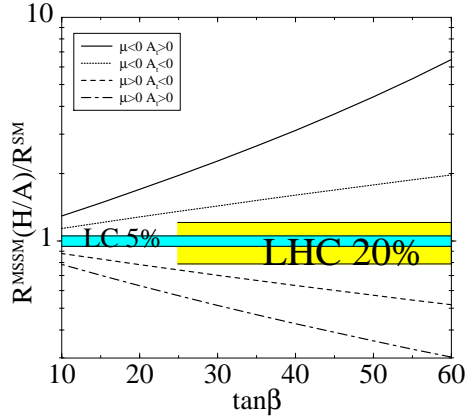}}}
\caption{Deviation of $\rmssm(H/A)$ with respect to the SM value, as a
  function of \tb\  for various choices
  of the SUSY parameters. The shaded regions are as in
  Fig.~\ref{fig:rmssmh}.}
\label{fig:rmssmH}
\end{figure}

We have found also the regions in the $(\tb,\ma)$ plane in which each
experiment can be sensitive to the SUSY nature of the lightest Higgs
boson. We show these regions in Fig.~\ref{fig:exclregion}a for a $5.4\%$
accuracy measurement, and in Fig.~\ref{fig:exclregion}b for a $21\%$ one.
We see that with a 5.4\% measurement one can have sensitivity to SUSY
for \ma\  up to $\sim 1.8\TeV$ in the most favorable scenario. In
less-favored scenarios the sensitivity is kept up to   
$\ma\sim800\GeV$, but  there exists also large regions where one is
sensitive to SUSY only up to $\ma\sim500\GeV$. 
However, all these masses are
well above the threshold production of the heavy Higgs particles for a
$500\GeV$ LC. We stress once 
again that these conclusions are independent of the scale of the SUSY
masses. As long as a $21\%$ accuracy is concerned, feasible e.g.\ at
the LHC, the regions of sensitivity are of course much smaller
(Fig.~\ref{fig:exclregion}b). In this case one can probe the SUSY
nature of the Higgs boson only if $A^0$ is lighter that
$\sim900\GeV$. This means that the heavier MSSM Higgs bosons $H^0$, $A^0$ and
$H^\pm$ will also be 
produced at high rates at the LHC. Then, it would be more useful to
move our attention to $\rmssm(H/A)$. We have checked that this 
quantity is very insensitive to $\tan\alpha$, and so to \ma. Its
numerical value is very close for both types of heavy neutral Higgs
bosons. A deviation of 21\% with respect to the SM value
is guaranteed for any scenario with $\tb\gsim20$ as shown in 
Fig.~\ref{fig:rmssmH}; hence, the SUSY
nature of the Higgs sector can be determined with a  
moderate-precision measurement. Notice that for the LHC models can be
only distinguished for $\tb > 25$ but for th LC cover the entire $\tb$ range.

To summarize, we have proposed the observable $R=\brHbb/\brHtt$ to
discriminate between SUSY and non-SUSY Higgs models. This observable
suffers  only little from systematic uncertainties, 
and is a theoretically
\textit{clean} observable.  In the MSSM, $R$ is affected by 
 quantum contributions that do not decouple 
even in the heavy SUSY limit. By assuming a $\pm5.4\%$
measurement of this ratio for the lightest Higgs boson, to be reached at a
$500\GeV$ LC, one is sensitive to the SUSY nature of the lightest Higgs
boson $h^0$ for values of the $A^0$
mass up to $1.8\TeV$.  A less precise measurement at
$\pm21\%$ accuracy, feasible at the LHC,
is sensitive to SUSY only if
$\ma<900\GeV$. In this latter case the measurement of $R$
for the heavy 
Higgs bosons $A^0$ and $H^0$ is possible and  can give clear evidence
for, or against, the SUSY nature of the Higgs bosons. 
Further confirmation can be obtained by correlating these 
measurements with the 
production cross-section of charged Higgs bosons~\cite{Belaetal}.
Further simulation
analysis of the expected experimental determination are highly desirable.


\bigskip
 
\noindent 
\small
\begin{thebibliography}{10} 

\bibitem{ours}
J.~Guasch, W.~Hollik, S.~Pe{\~n}aranda,
Phys.\ Lett.\ B {\bf 515} (2001) 367, hep-ph/0106027.
 
\bibitem{eff} 
M.~Carena, D.~Garcia, U.~Nierste, C.E.M.~Wagner, Nucl.\ Phys.\  \textbf{B577} 
(2000) 88, 
hep-ph/9912516. 

\bibitem{Guasch-Spira}
J.~Guasch, P.~H\"afliger, M.~Spira, hep-ph/0305101.
 
\bibitem{CMW} 
M. Carena, S. Mrenna, C.E.M. Wagner, Phys. Rev. \textbf{D60} (1999) 
075010, hep-ph/9808312; 
\textit{ibid.} {\bf D62} (2000) 055008, hep-ph/9907422;
D.~M.~Pierce, J.~A.~Bagger, K.~Matchev, R.~Zhang,
Nucl.\ Phys.\  {\bf B491} (1997) 3, hep-ph/9606211;
M.~Carena, M.~Olechowski, S.~Pokorski, C.E.M.~Wagner, Nucl.\ 
Phys.\ \textbf{B426} (1994) 269, hep-ph/9402253; 
L.J.~Hall, R.~Rattazzi, U.~Sarid, Phys.\ Rev.\ \textbf{D50} (1994) 
7048, hep-ph/9306309. 

\bibitem{PDG} D.~E.~Groom {\it et al.}  [Particle Data Group Collaboration],
Eur.\ Phys.\ J.\  {\bf C15} (2000) 1.

\bibitem{FeynHiggsFast}
S.~Heinemeyer, W.~Hollik, G.~Weiglein,
hep-ph/0002213.

\bibitem{Zeppenfeld}
D.~Zeppenfeld, R.~Kinnunen, A.~Nikitenko, E.~Richter-Was,
Phys.\ Rev.\  {\bf D62} (2000) 013009, hep-ph/0002036.

\bibitem{Gianotti}
F.~Gianotti, M.~Pepe-Altarelli,
Nucl.\ Phys.\ Proc.\ Suppl.\  {\bf 89} (2000) 177, hep-ex/0006016.


\bibitem{TESLATDR} \textit{2nd Joint ECFA/DESY Study on Physics and
                Detectors for a Linear Electron-Positron Collider},
                \texttt{http://www.desy.de/conferences/ecfa-desy-lc98.html};
                \textit{TESLA Technical Design Report} \texttt{DESY
                2001-011}, Part III: \textit{Physics at an $e^+e^-$ Linear
                Collider}, R. Heuer, D. Miller, F. Richard, P. Zerwas
                Editors, hep-ph/0106315, \texttt{http://tesla.desy.de/}.

\bibitem{Belaetal} 
A.~Belyaev, D.~Garcia, J.~Guasch, J.~Sol{\`a},
Phys.\ Rev.\ D {\bf 65} (2002) 031701, hep-ph/0105053; 
JHEP {\bf 0206} (2002) 059, hep-ph/0203031.


\end{thebibliography}
\end{document}